\documentclass{elsart}
\usepackage{epsfig}
\journal{New Astronomy}
\begin{document}
\begin{frontmatter}

\title{The contribution of AGN to the X--ray background: 
       the effect of iron features}

\author[astfi,ira]{R. Gilli},
\author[ossbo]{A. Comastri},
\author[astbo,ira]{G. Brunetti},
\author[astbo,ira]{G. Setti}

\address[astfi]{Dipartimento di Astronomia, Universit\`a di Firenze, Largo
E. Fermi 5, I--50125 Firenze, Italy}
\address[ira]{Istituto di Radioastronomia del CNR, Via Gobetti 101,
I--40129 Bologna, Italy}
\address[ossbo]{Osservatorio Astronomico di Bologna, Via Ranzani 1,
I--40127 Bologna, Italy}
\address[astbo]{Dipartimento di Astronomia, Universit\`a di Bologna, Via
Ranzani 1, I--40127 Bologna, Italy}

\begin{abstract}
The contribution of the iron emission line, commonly detected 
in the X--ray spectra of Seyfert (Sey) galaxies, to the cosmic X--ray
background
(XRB) spectrum is evaluated in the framework of the XRB synthesis
models based on AGN unification schemes. 
To derive the mean line properties, we have carried out 
a search in the literature covering a sample of about 70 AGN.   
When adopting line parameters in agreement with the observations,
it turns out that the maximum contribution of the iron line to the XRB is 
less than 7\% at a few keV.
This is still below the present uncertainties in the XRB spectrum
measurements. 
\end{abstract}
 
\end{frontmatter}


\section{Introduction}

The observations performed by the HEAO--1, ROSAT and ASCA satellites
have allowed a good description of the X--ray background spectrum over a
wide energy range  
from 0.1 to $\sim 400$ keV.
The intensity above $\sim1$ keV has an extragalactic origin, resulting
from the summed emission of unresolved sources over all cosmic epochs.   
The recent ROSAT deep survey in the Lockman hole has shown that in the soft
0.5--2.0 keV band nearly $70\%$ of the XRB is already resolved into discrete
sources
at a limiting flux of $\sim 10^{-15}$ erg cm$^{-2}$ s$^{-1}$ (Hasinger et
al. 1998). 
The resolved fraction in the 2--10 keV band is much lower:
from a medium sensitivity survey with
the ASCA GIS about $27\%$ of the XRB measured by HEAO--1 (Gruber 1992) has
been resolved at a limiting flux of
$\sim 6\times10^{-14}$ erg cm$^{-2}$ s$^{-1}$ (Cagnoni, Della Ceca \& 
Maccacaro 1998).
While nearly $80\%$ of the ROSAT sources have been optically 
identified as active galactic nuclei (AGN; Schmidt et al. 1998), 
the optical counterparts of the ASCA sample are not fully available yet. 

On the other hand it is well known that the 2--20 keV XRB spectrum, well
approximated by a flat power
law with energy spectral index $\alpha\sim 0.4$, can not be simply reproduced
by the emission of quasars and Sey 1 galaxies, which have
steeper 
($\alpha\sim0.8-0.9$) hard X--ray spectra (Comastri et al. 1992; 
Nandra \& Pounds 1994). 
This apparent contradiction, known as the ``spectral paradox'' (Boldt
1987),
has been circumvented by Setti \& Woltjer (1989) who demonstrated that
the main characteristics of the XRB could be reproduced by summing the 
contributions from
unabsorbed (type 1) and absorbed (type 2) AGN according to the population
ratios predicted by the unification models (Antonucci \& Miller 1985; Barthel
1989). Comastri et al. (1995) have since worked out a detailed model which 
provides a
good fit to the XRB in the broad 3--100 keV band and is consistent with all
known
X--ray properties of AGN classes (see also Madau et al. 1994). Other AGN
models
of the XRB do not meet observed constraints on the X--ray spectra and/or
source
counts (see Setti \& Comastri 1996 for a review).

However, it has been pointed out (Matt \& Fabian 1994) that strong iron 
features, such as the line at $\sim6.4$ keV and 
related absorption edge commonly found in Seyferts spectra 
(Nandra \& Pounds 1994), 
may lead AGN synthesis models to predict a 
detectable feature in the XRB spectrum at a few keV.
Since recent ASCA observations (Gendreau et al. 1995) have
shown that the 1.7--10 keV XRB spectral shape is rather smooth and 
featureless, apart from a narrow bump at $\sim2$ keV due to the gold 
edge in the instrumental response,
Di Matteo \& Fabian (1997) have
proposed that a new and as yet undetected population of
advection--dominated
sources with flat X--ray spectra is required to fit the XRB spectrum.

In order to further investigate whether the iron structure in the XRB
spectral profile should have been detected, we have reconsidered 
the AGN synthesis 
model of the XRB described by Comastri et al. (1995) adding the iron lines
to the input spectra.

Throughout this paper the values $q_{0}=0$ and $H_{0}=50$ km s$^{-1}$
Mpc$^{-1}$ have been assumed.

\section{The iron line}

The X--ray spectra of Seyfert galaxies commonly show radiation excesses in
the 5--7 keV band with respect to a power law continuum. These are usually 
interpreted as K$\alpha$ iron emission lines from regions reprocessing the 
nuclear radiation.

The iron line properties strongly depend on
the geometry and dynamics of the reprocessing medium. From 
the energy of the line
peak $E_{\rm K\alpha}$, the width $\sigma_{\rm K\alpha}$ and the
equivalent width $W_{\rm K\alpha}$it is possible to identify the emission
line regions. In most cases it is found
that $E_{\rm K\alpha}\simeq6.4$ keV, in agreement with
fluorescence emission from iron in the ionization range Fe {\sc i}--Fe 
{\sc xvi}, which is thought to be distributed into a cold
($T<10^6$ K) accretion disc
as well as into a surrounding molecular dusty torus.

Since the disc and torus components may contribute differently to the
observed properties, the profiles and intensities of the iron lines are 
expected to be complex, depending also on the inclination with respect to
the line of sight.

The properties of the iron line emitted by a cold accretion disc
have been calculated in a number of
papers (George \& Fabian 1991; Matt et al. 1991).
When relativistic effects are considered (Matt et al. 1992), 
the calculated line profile is broad
($\sigma_{\rm K\alpha}\geq0.2$ keV),
asymmetric and double horned, with a Doppler boosted blue horn. The line
equivalent width is predicted to be at most $\sim200$ eV. Broad asymmetric
line profiles have been detected by ASCA in several Sey 1 galaxies
(Tanaka et al. 1995; Mushotzky et al. 1995; Nandra et al. 1997a)
in agreement with
relativistic disc models. 

Ghisellini, Haardt \& Matt (1994) have performed a quantitative analysis
of the fluorescence line emission from a specific torus configuration
under the assumption
that the disc and the torus are coplanar. In the case of type 1 AGN,
viewed along lines
of sight free from obscuration, the iron line emitted by a 
dense torus ($N_{\rm H} > 5\times10^{23}$ cm$^{-2}$) may
contribute substantially to the overall line strength (up to 
$W_{\rm K\alpha}\sim 90$ eV).
On the other hand,
since in type 2 AGN
the radiation from the central source is intercepted by the torus, the
equivalent width of the torus iron line may become very
large (up to a few keV) for tori that are optically thick to Compton
scattering (equatorial $N_{\rm H} >$ a few $\times 10^{24}$ cm$^{-2}$).
It is important to note that the iron line emitted by the torus should have
roughly a narrow ($\sigma_{\rm K\alpha}\leq0.1$ keV) gaussian profile.
The iron line equivalent widths and profiles of several Sey 2s are in 
good agreement
with the torus picture (Smith \& Done 1996; Weaver et al. 1996).
Very strong iron lines associated to flat
reflection continua are found in a dozen of type 2 objects, most
likely Compton thick AGN (Maiolino et al. 1998 and
references therein).

In a few objects, mainly Sey 2 galaxies, the iron emission
is due to a combination of neutral K$\alpha$ at 6.4 keV
plus lines from highly ionized iron (Fe {\sc xxv}, Fe {\sc xxvi})
at 6.68 keV and 6.96 keV (Ueno et al. 1994; Turner et al. 1997a,b).
These lines are thought to be produced in a `warm mirror' (Matt, Brandt \&
Fabian 1996),
which may also be responsible for the scattering of the broad
optical lines observed in polarized light in type 2 objects.

The iron lines (as well as
the reflection humps above $\sim 10$ keV) are not a common feature of the 
quasars spectra
(Lawson \& Turner 1997). Iwasawa \& Taniguchi (1993) have suggested that  
an `X--ray Baldwin effect' may hold, whereby the equivalent width of the  
lines decreases with
increasing luminosity. This has been confirmed by Nandra et al. (1997b),  
who also find a change in the line profile with luminosity and
propose a model where the accretion disc becomes more ionized
as a function of the accretion rate.  

In order to estimate the mean properties of the iron line we have
collected literature data based on ASCA and GINGA measurements.
We have also included recent data from BeppoSAX observations of
several Sey 2 galaxies
(Salvati et al. 1997; Maiolino et al. 1998).
The sample consists of 29 Sey 1 galaxies, 36 Sey 2s
and 6 Narrow Emission Line Galaxies (NELGs).
Among the AGN in the sample, 58 objects have
been observed by ASCA and 41 by GINGA, with 34 objects in both samples.
The $W_{\rm K\alpha}$ and $N_{\rm H}$ values for the AGN in our
samples are shown in Table~1. 

\begin{table*}
\caption{Line equivalent widths and column densities for the ASCA and
GINGA samples}
\begin{tabular}{lllc|llc}
\multicolumn{7}{c}{Type 1 AGN}\\
\hline \hline
Name& & ASCA& & & GINGA& \\
& $W_{\rm K\alpha}$(eV)& log[$N_{\rm H}$(cm$^{-2}$)]& Ref.& 
$W_{\rm K\alpha}$(eV)& log[$N_{\rm H}$(cm$^{-2}$)]& Ref.\\ 
\hline
NGC 3227& $206\pm22$& $20.98\pm0.04$& 1,2&
$150\pm30$& $21.32^{+0.15}_{-0.24}$& 3\\
NGC 3516& $126\pm18$& $<19.90$& 1,2& $110\pm90$&
$22.76^{+0.03}_{-0.04}$& 3\\
NGC 3783& $132\pm18$& $20.66^{+0.04}_{-0.05}$& 1,2& $180\pm40$&
$22.34\pm0.03$& 3\\
NGC 4051& $400^{+97}_{-78}$& $<19.78$& 2& $230\pm30$& $<21.70$& 4\\
NGC 4151& $307^{+148}_{-88}$& $23.05^{+0.08}_{-0.07}$& 5& $160\pm20$&
$23.05^{+0.03}_{-0.02}$& 4\\
NGC 4593& $164\pm35$& $20.30^{+0.06}_{-0.10}$& 1,2& $154\pm27$&
$<21.30$&
3,4\\
NGC 5548& $244\pm34$& $<19.30$& 1,2& $112\pm13$&
$<21.2$& 3,4\\
NGC 6814& $430^{+479}_{-182}$& $<20.76$& 2& &
 & \\
NGC 7213& & & & $130\pm40$& $21.46^{+0.16}_{-0.26}$& 3\\
NGC 7469& $147\pm30$& $<19.60$& 2,6&$98\pm70$& $<21.11$& 3,7\\
Mrk 290& $448\pm56$& $<19.78$& 2,8& & & \\
Mrk 335& $167\pm56$& $<19.95$& 1,2& $155\pm70$&
$<21.62$& 3,9\\
Mrk 509& $210^{+190}_{-130}$& $20.32^{+0.06}_{-0.07}$& 1,2& $60\pm30$&
$<20.78$& 3,4\\
Mrk 766& $115\pm24$& $<20.28$& 1,10& & & \\
Mrk 841& $130\pm52$& $<21.00$&1,2& $447\pm184$& $21.73^{+0.34}_{-1.26}$&
3\\
Mrk 1040& $550^{+152}_{-152}$& $21.56^{+0.05}_{-0.07}$& 11& & & \\
3C 111& $40-195$& $21.81\pm0.01$& 12& $<80$& $22.26\pm0.05$& 3\\
3C 120& $399^{+81}_{-74}$& $21.22\pm0.01$& 13& & & \\  
3C 382& $900^{+632}_{-188}$& $20.04^{+0.21}_{-0.26}$& 2& $200\pm60$&
$<21.79$& 3,4\\
3C 390.3& $251\pm40$& $20.86\pm0.04$& 2,14& $43\pm20$& $<20.90$& 3,4\\
3C 445& $268^{+165}_{-72}$& $22.76^{+0.19}_{-0.16}$& 15&  & & \\
MCG--2-58-22& $320\pm60$& $<20.56$& 1,2,16& $150\pm50$&
$<21.89$& 3\\
MCG--6-30-15& $338\pm64$& $20.23\pm0.05$& 1,17& $135\pm14$&
$21.66^{+0.07}_{-0.08}$& 3,18\\
IRAS 15091--2107& & & & $190\pm40$& $<22.00$& 4\\
ESO 141--G55& $140\pm42$& $20.72^{+0.07}_{-0.06}$& 2& & & \\
IC 4329A& $132\pm18$& $21.49\pm0.01$& 1,2,19& $110\pm20$&
$21.68^{+0.04}_{-0.05}$& 3\\
Fairall 9& $369\pm54$& $20.08^{+0.15}_{-0.18}$& 1,2& $120\pm70$&
$<21.15$& 3\\
Akn 120& & & & $100\pm50$& $<21.40$& 3,9\\
H2106--099& & & & $<180$& $<21.84$& 3\cr  \hline
\end{tabular}
\end{table*}

\setcounter{table}{0}

\begin{table*}
\caption{-Continued}
\begin{tabular}{lllc|llc}
\multicolumn{7}{c}{Type 2 AGN}\\
\hline \hline
Name& & ASCA& & & GINGA& \\
& $W_{\rm K\alpha}$(eV)& log[$N_{\rm H}$(cm$^{-2}$)]& Ref.&
$W_{\rm K\alpha}$(eV)& log[$N_{\rm H}$(cm$^{-2}$)]& Ref.\\
\hline
NGC 526A& $175^{+62}_{-53}$& $22.18^{+0.03}_{-0.05}$& 20& $212\pm87$&
$22.05^{+0.42}_{-0.18}$& 3,21\\
NGC 1667& $<3000$& $<22.51$& 20&$<390$& $<23.00$& 22\\
NGC 1672& $<1600$& $<22.48$& 23& $<150$& $<22.50$& 22\\
NGC 1808& $336^{+453}_{-336}$& $<22.96$& 20& & & \\ 
NGC 2110& $206^{+108}_{-66}$& $22.45\pm0.03$& 20& $154\pm26$&
$22.38^{+0.04}_{-0.11}$& 3,21\\
NGC 2992& $555\pm60$& $21.84^{+0.11}_{-0.08}$& 2,20,24& $328\pm42$&
$22.21^{+0.10}_{-0.28}$&
3,21\\
NGC 3281& $751^{+232}_{-162}$& $23.90^{+0.06}_{-0.05}$& 25& & & \\ 
NGC 4258& $250\pm61$& $23.18^{+0.03}_{-0.04}$& 26& & & \\
NGC 4388& $487\pm48$& $23.58^{+0.02}_{-0.03}$& 23,27& & & \\
NGC 4507& $152\pm22$& $23.57\pm0.02$& 20,23,28& $469\pm43$&
$23.61^{+0.08}_{-0.06}$& 21,22\\
NGC 5252& $98\pm38$& $22.64\pm0.03$& 20,29& & & \\
NGC 5506& &  & & $196\pm12$& $22.54\pm0.02$& 3,21\\
NGC 5674& &  & & $<140$& $22.84^{+0.09}_{-0.12}$& 21\\
NGC 6251& $228^{+219}_{-188}$& $21.09^{+0.11}_{-0.14}$& 20& & & \\
NGC 7172& $72\pm16$& $22.90\pm0.01$& 20,23& $71\pm18$&
$23.05\pm0.02$&
3,21,22\\
NGC 7314& $550\pm174$& $22.06^{+0.02}_{-0.06}$& 20,30& $104\pm37$&
$21.74^{+0.09}_{-0.11}$&
3,21\\
NGC 7319& $624^{+137}_{-159}$& $23.52^{+0.10}_{-0.22}$& 23& & & \\
NGC 7582& $281^{+210}_{-153}$& $22.87^{+0.08}_{-0.06}$& 20& & & \\
Mrk 3& $862^{+283}_{-183}$& $23.67\pm0.06$& 31,32& $550\pm67$&
$23.85\pm0.03$& 21,22\\
Mrk 348& & & & $177\pm49$& $23.03\pm0.08$& 21,22\\
Mrk 463E& $429^{+349}_{-320}$& $<22.85$& 20,23& & & \\
Mrk 477& $490^{+152}_{-121}$& $<23.21$& 23& & & \\
Mrk 1210& $830^{+272}_{-224}$& $23.08^{+0.15}_{-0.22}$& 23& & & \\
MCG--5-23-16& $455^{+194}_{-131}$& $22.20^{+0.02}_{-0.02}$& 20&
$336\pm31$& $22.25^{+0.04}_{-0.05}$& 3,21\\
IRAS 04575--7537& $158^{+28}_{-29}$& $22.09\pm0.01$& 33&
$260^{+69}_{-66}$& $22.07^{+0.21}_{-0.10}$& 21\\
IRAS 18325--5926& $580^{+240}_{-150}$& $22.12\pm0.04$& 34& $198\pm33$&
$22.21^{+0.10}_{-0.05}$& 21,22\\
IRAS 20460+1925& $260^{+88}_{-83}$& $22.40^{+0.04}_{-0.03}$& 35& & &
\\
ESO 103--G35& $505^{+265}_{-192}$& $23.20^{+0.88}_{-0.68}$& 20&
$370\pm55$& $23.38\pm0.03$& 21\\
IC 5063& $80^{+30}_{-25}$& $23.38\pm0.02$& 23& $272\pm57$&
$23.37^{+0.03}_{-0.03}$&
21,36\\
PKS B1319--164& $327^{+74}_{-64}$& $23.61\pm0.03$& 23& & & \cr
\hline
\end{tabular}
\end{table*}

\setcounter{table}{0}

\begin{table*}
\caption{-Continued}
\begin{tabular}{lllc|llc}
\multicolumn{7}{c}{Compton thick type 2s}\\
\hline \hline
Name& & ASCA& & & GINGA& \\
& $W_{\rm K\alpha}$(eV)& log[$N_{\rm H}$(cm$^{-2}$)]& Ref.&
$W_{\rm K\alpha}$(eV)& log[$N_{\rm H}$(cm$^{-2}$)]& Ref.\\
\hline
NGC 1068& $1321\pm226$& $>24.18$& 20,37& $2350^{+933}_{-442}$& $>24.18$&
21\\
NGC 4945& $1450^{+530}_{-420}$& $>24.18$& 20& $1500\pm182$&
$24.76^{+0.02}_{-0.03}$& 38\\
NGC 4968& $1180^{+4420}_{-827}$& $>24.18$& 20& & & \\ 
NGC 5135& $4690^{+1580}_{-1460}$& $>24.18$& 20& & & \\
NGC 6240& $1580\pm230$& $>24.18$& 20,39& & & \\
CIRCINUS& $2310^{+73}_{-158}$& $>24.18$& 40& & & \\ 
\cline{1-4}
& & BeppoSAX& & & & \\ \cline{1-4}
NGC 1386& $7600^{+5400}_{-3000}$& $>24.18$& 41& & & \\
NGC 2273& $2490^{+485}_{-412}$& $>25$& 41& & & \\
NGC 3393& $1890^{+2300}_{-727}$& $>25$& 41& & & \\  
NGC 4939& $480^{+255}_{-127}$& $>25$& 41& & & \\
NGC 5463& $1900^{+848}_{-424}$& $>25$& 41& & & \\
NGC 7674& $900^{+285}_{-181}$& $>24.18$& 42& & & \cr \hline
\end{tabular}

Errors are 68\% confidence limits. For the objects with more than one
reference the quoted values derive from weighted means.\\
The equivalent widths values measured by ASCA derive from broad line fits.
Type 1 AGN collect objects with optical spectrum from Sey 1.0 to 1.5;
Type 2 AGN collect Seyferts from 1.8 to 2.0 and NELGs.\\
References. -- (1) Nandra et al. 1997a; (2) Reynolds 1997; (3)
Nandra \& Pounds 1994; (4) Awaki et al. 1991; (5) Yaqoob et al. 1995; 
(6) Guainazzi et al. 1994; (7) Piro et al. 1990; (8) Turner et
al. 1996; (9) Iwasawa \& Taniguchi 1993; (10) Leighly et al. 1996;
(11)
Reynolds et al. 1995; (12) Reynolds et al. 1998; (13) Grandi et al.
1997;
(14) Eracleous et al. 1996; (15) Sambruna et al. 1998; (16) Weaver et
al.
1995; (17) Tanaka et al. 1995; (18) Awaki 1991; (19) Cappi et al. 1996a;
(20) Turner et
al.
1997a; (21) Smith \& Done 1996; (22) Awaki \& Koyama 1993; (23) Ueno 
1997; (24) Weaver et al. 1996; (25) Bassani et al. 1998; (26) Makishima
et
al. 1994; (27) Iwasawa et al. 1997; (28) Comastri et al. 1998; (29)
Cappi
et al. 1996b; (30) Yaqoob et al. 1996; (31) Iwasawa et al. 1994; 
(32) Griffiths et al. 1998; 
(33) Vignali et al. 1998; (34) Iwasawa et al. 1996; (35) Ogasaka et al.
1997;
(36) Koyama et al. 1992; (37) Ueno et al. 1994; (38) Iwasawa et al.
1993; (39) Iwasawa \& Comastri 1998; (40) Matt et al.
1996; (41) Maiolino et al. 1998; (42) Malaguti et al. 1998.
\end{table*}

\subsection{ASCA data}

\begin{table*}
\caption{Mean iron line parameters from the ASCA and the GINGA samples.}
\begin{tabular}{@{}c@{~~}l@{~}c@{~}c@{~}c@{~}c@{~}c@{~}c}
\hline \hline AGN type& ~Sample&
\multicolumn{6}{c}{Mean iron line parameters}\\ & &
\multicolumn{2}{c}{$E_{\rm K\alpha}$(keV)}&
\multicolumn{2}{c}{$\sigma_{\rm K\alpha}$(keV)}&
\multicolumn{2}{c}{$W_{\rm K\alpha}$(eV)}\\ & &
$m^a$&
$ML^b$&
$m$&
$ML$&
$m$&
$ML$\\
\hline & & & & & \\ type 1& ASCA&
$6.36\pm0.13^{(23)}$&
$6.36^{+0.03}_{-0.02}$&
$0.51\pm0.37^{(20)}$& $0.40\pm0.05$& $287\pm180^{(24)}$& $229\pm23$\\
& GINGA&
$6.50\pm0.30^{(16)}$& $6.38^{+0.04}_{-0.05}$& & &  
$152\pm83^{(20)}$& $135\pm 13$\\
& & & & & \\ type 2& ASCA& $6.28\pm0.32^{(22)}$&
$6.38^{+0.02}_{-0.01}$&
$0.34\pm0.27^{(14)}$& $0.27\pm0.06$& $390\pm231^{(24)}$& $322\pm43$\\ 
& GINGA& $6.45\pm0.18^{(11)}$& $6.46\pm0.04$& & &
$264\pm135^{(14)}$& $259\pm35$\\ & & & & &\\
Compton& ASCA& $6.40\pm0.11^{(5)}$& &
$0.20\pm0.07^{(4)}$& &
$2089\pm1334^{(6)}$& \\
thick& BeppoSAX& & & & & $2543\pm2583^{(6)}$& \cr \hline
\end{tabular}

Errors are 68\% confidence limits. The number of objects
are in small brackets.\\
$^a$ Mean and standard deviation.\\
$^b$ Maximum Likelihood estimate.\\
\end{table*}
Most of the ASCA data are taken from the samples
of Nandra et al. (1997a), Reynolds (1997) and Turner et al. (1997a), 
plus other sources analyzed in various papers.
The AGN have been
divided into two classes: type 1 objects, grouping Sey 1.0--1.2--1.5, and
type 2 objects, grouping Sey 1.8--1.9--2.0 as well as NELGs because
of their substantial intrinsic absorption.
Furthermore, we have subdivided type 2s into
Compton thin and Compton thick AGN.
The average line parameters derived from the sample are reported in
Table~2. In this and in the following Tables the mean $W_{\rm
K\alpha}$
values are calculated without considering those objects where only upper
limits are available (see Table~1). We note here that the derived mean 
values are generally consistent with the median values within 20\%.
Whenever
dealing with good statistical samples, we have estimated the mean and the
intrinsic dispersion of the parent population with the Maximum Likelihood
algorithm (henceforth ML, Maccacaro et al. 1988), assuming a gaussian
intrinsic distribution of the parameters.
Confidence intervals are given at the 68\% confidence level for one
interesting parameter.

\begin{figure}
\epsfig{file=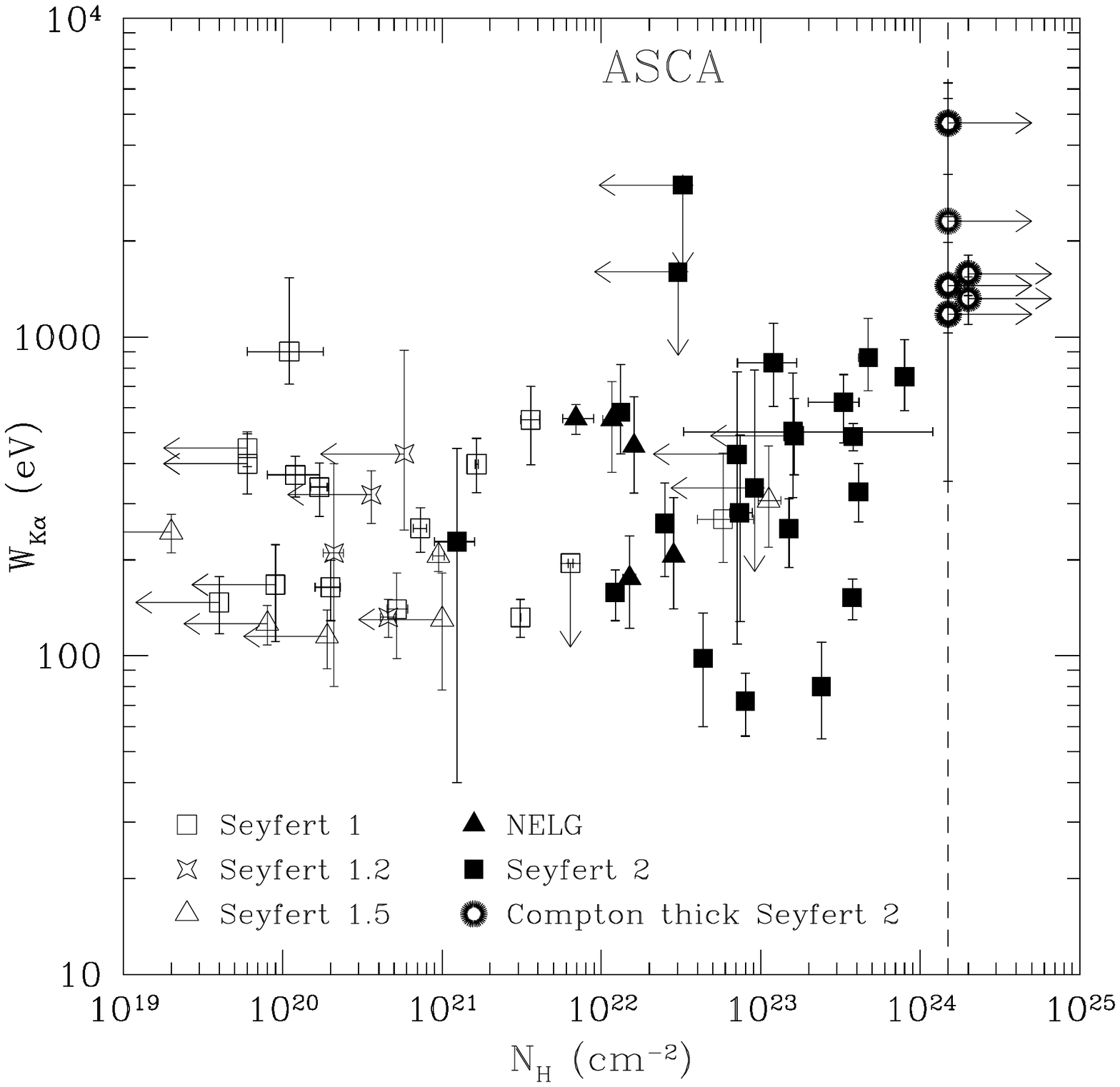, height=9.cm, angle=0, silent=}
\caption{The line equivalent width $W_{\rm K\alpha}$ plotted against the
absorbing column $N_{\rm H}$ for the ASCA sample. From ASCA observations
only a lower limit to the column density of Compton thick AGN, which
have
been splitted for clarity, can be set (see text).}
\end{figure}

For both type 1 and type 2 objects the average energy of the line
peak is close to 6.4
keV, as expected from emission by fluorescent cold iron.
Since ASCA has good spectral resolution
($\Delta E=120$ eV at 6 keV) the
line profile can be carefully fitted allowing a precise measure of the
line width.
Type 1 objects show broad lines with a ML mean
$\langle{\sigma}_{\rm K\alpha}\rangle=0.40\pm0.05$ keV in
agreement with the
value quoted by Nandra et al. (1997a), suggesting that 
the iron emission is dominated by the 
relativistic accretion disc component.
Compton thin type 2 objects have a ML mean width $\langle{\sigma}_{\rm
K\alpha}\rangle=0.27 \pm 0.06$ keV, mostly due to the broad
lines of NELGs, suggesting that a disc line component may still be
significant (Weaver \& Reynolds 1998).

Two opposite effects are relevant in the ASCA measurements of the iron
line equivalent widths:
good energy resolution allows a careful fit to the line
flux above the underlying continuum, but the limited energy band
covered by the satellite (0.5--10 keV) and especially the low effective
area above 7 keV prevent a detailed modeling of the continuum. 
The $W_{\rm K\alpha}$ measurements spread is
clearly visible in Fig.~1, where it is also evident that type 1 and 2
AGN are divided into two groups according to their X--ray absorption,
the dividing column
density being $N_{\rm H}\sim 10^{21.5-22}$ cm$^{-2}$.

If Compton thick AGN are taken into account,
a correlation between $W_{\rm K\alpha}$ and $N_{\rm H}$ may be recognized
for type 2 objects. Nevertheless, the data dispersion is very high and
a selection effect may be present, the strongest lines
being more easily detected.
This trend is predicted by theoretical models 
(Ghisellini et al. 1994; Awaki 1991) whereby, given the average inclination of
type 2 objects with respect to the line of sight, 
the observed line intensity produced by reflection and transmission
in the torus increases with equatorial $N_{\rm H}$, while the continuum flux
is reduced, leading to an enhancement of the equivalent width at the
highest column densities.
If Compton thick AGN are not considered, there is still some evidence of this
correlation in the last $N_{\rm H}$ bin (see also Table 4), but the number of 
objects is small and the sample incomplete. 

The mean equivalent widths of type 1s and 2s, estimated with the ML
method, are $\langle W_{\rm K\alpha}\rangle=229\pm23$ eV and
$\langle W_{\rm K\alpha}\rangle=322\pm43$ eV, respectively. The ML average
value quoted by 
Nandra et al. (1997a) for their type 1s sample is $160\pm30$ eV 
(lines fitted with broad gaussians), which is lower than our result, but 
consistent with it within $2\sigma$.
The average ML equivalent width in the sample of Turner et al.
(1997a), which however includes
several Compton thick objects, is $\langle W_{\rm
K\alpha}\rangle=452^{+184}_{-144}$ eV, statistically consistent with 
our estimate. 

When considering the whole ASCA sample, except Compton thick AGN, the
average equivalent width is $W_{\rm
K\alpha}=340\pm212$ eV for the simple mean and
$\langle W_{\rm K\alpha}\rangle=265\pm24$ eV for the ML estimate.

\subsection{GINGA data}

Most of the data from GINGA are taken from Nandra \& Pounds (1994) and Smith \&
Done (1996)
who analyzed a sample of type 1 and type 2 sources, respectively.
Since GINGA detectors are not sensitive below $\sim$2 keV,
only an upper limit to the intrinsic absorption of many type 1s can be
set.
The high absorption revealed in NGC 3516, NGC 3783 and MCG--6-30-15, in
disagreement with ASCA data (Table~1), is due to the presence of a  warm
absorber causing a turnover in the spectra of these sources at the low
energy end
of the GINGA band that mimics a cold absorption in the GINGA data 
(e.g., Mathur, Wilkes \& Aldcroft 1997).                          

As for the ASCA data, the mean energy of the line peak $\langle E_{\rm
K\alpha}\rangle$ is fully consistent with 6.4 keV (Table~2),
but no information is available on the line's widths due to the poor energy
resolution ($\Delta E=1.2$ keV at 6 keV) of the GINGA instruments, which
does not allow a careful sampling of the line profile. Spectral fits have
always been obtained by adopting gaussian lines of width 0.1 or 0.05 keV, much
narrower than the instrumental response.

\begin{figure}
\epsfig{file=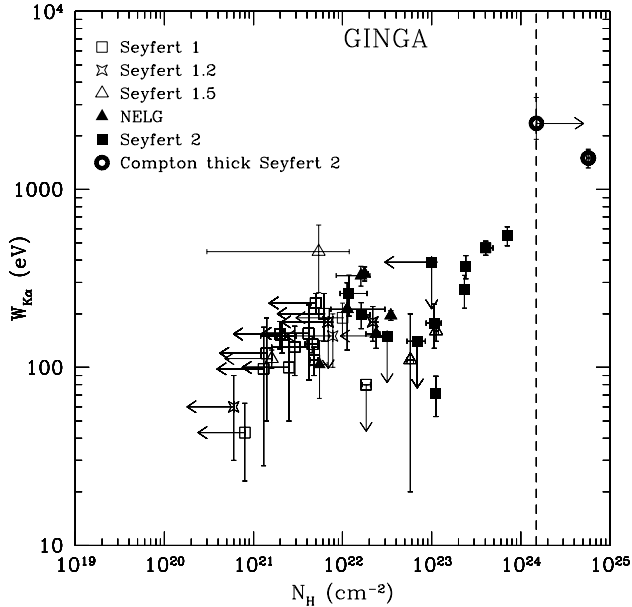, height=9.cm, angle=0, silent=}
\caption{The line equivalent width $W_{\rm K\alpha}$ plotted against the
absorbing column $N_{\rm H}$ for the GINGA sample.}
\end{figure}

The spread in the  $W_{\rm K\alpha}$ measurements is lower than 
in the ASCA sample (Fig.~2). 
Notice that
for the source with $N_{\rm H}\simeq10^{24.6}$ cm$^{-2}$, namely NGC4945, 
the detection by GINGA of the nuclear radiation above $\sim10$ keV,
has led to 
a direct measurements of the torus column density (Iwasawa et al. 1993).
As expected from the theoretical models,
$W_{\rm K\alpha}$ and $N_{\rm H}$ seem to be correlated (Fig.~2).
We feel that this conclusion should be taken with due care because, like
for the ASCA sample, the number of the most absorbed sources ($N_{\rm
H}>2-3\times 10^{23}$ cm$^{-2}$) is very small and the sample may be biased.

The mean equivalent widths for our GINGA sample are given in Table~2. 
From the ML analysis   
$\langle W_{\rm K\alpha} \rangle = 135\pm13$ for type 1 AGN, in agreement 
with the value of $140\pm 20$ eV quoted by Nandra \& Pounds (1994), while
for type 2 AGN $\langle W_{\rm K\alpha} \rangle = 259\pm35$ eV,
which is higher, although statistically consistent,
than the value of $200\pm 40$ eV obtained by Smith \& Done (1996). 

The mean equivalent width for the whole sample is $W_{\rm
K\alpha}=200\pm118$ eV, for the mean and 
$\langle W_{\rm K\alpha} \rangle = 190\pm18$ eV for the ML estimate. 
\subsection{ASCA and GINGA common sample}

\begin{table} [b]
\caption{Results from the ASCA and GINGA common sample.}
\begin{tabular}{clcc}
\hline \hline AGN type& Common Sample&  
 \multicolumn{2}{c}{$W_{\rm K\alpha}$(eV)}\\ 
& & $m^a$& $ML^b$\\ \hline
& & & \\ type 1& ASCA& $267\pm186^{(17)}$& $208\pm20$\\ & GINGA&
$154\pm89^{(17)}$& $134\pm14$\\ & & & \\ 
type 2& ASCA& $363\pm254^{(12)}$& $279\pm60$\\ & GINGA&
$300\pm163^{(12)}$& $286\pm42$\cr \hline

\end{tabular}

Errors are 68\% confidence limits. The number of objects  
are in small brackets.\\
$^a$ Mean and standard deviation.\\
$^b$ Maximum Likelihood estimate.\\
\end{table}

To evaluate possible effects on the line equivalent width due to the
different instruments on board ASCA and GINGA, 
we consider here only the subsample of objects observed by both satellites. 
The results are shown in Table~3.
   
For type 1 objects the average
value $\langle W_{\rm K\alpha}\rangle$ derived from ASCA data is
significantly higher than that from GINGA, while for type 2 AGN the two
values are closer. 
It is suggested that the discrepancy for type 1 objects
could possibly be explained by the different spectral resolutions
of the 
two satellites: when the broad iron lines of type 1 AGN ($\sigma_{\rm
K\alpha}\sim0.4$ keV) are fitted with narrow
gaussians ($\sigma_{\rm K\alpha}\leq0.1$ keV), as
is the case for GINGA fits, it is likely that the flux of the line wings
is lost, resulting in an underestimate of the equivalent widths of the
lines; on the other hand, this effect should be less important for type 2
AGN, where the narrow line components emitted by the tori make a
substantial contribution to the measured line strengths.

As a consequence of the above results and considerations,
in our model computations we will only
consider $W_{\rm K\alpha}$ values based on ASCA measurements.  

\section{The Model}

The baseline model adopted to fit the XRB is that described in Comastri et
al. (1995), where we have introduced slight changes in
the parameter set
describing the AGN X--ray luminosity function (XLF) and evolution in
order
to
take into account the recent results obtained by Jones et al. (1997) and
Page et al. (1996).

Following Comastri et al. (1995), we assume a pure luminosity
evolution
$L_{x}(z)=L_{x}(0)\times(1+z)^{\beta}$, with $\beta=2.6$ and the XLF of
the form:

\begin{equation}
\rho(L,z=0)=K_{1}L_{44}^{-\gamma_{1}}\;\;{\rm for}\;L_{x}<L_{B}
\end{equation}
\begin{equation}
\rho(L,z=0)=K_{2}L_{44}^{-\gamma_{2}}\;\;{\rm for}\;L_{x}>L_{B}
\end{equation}

where $L_{44}$ is the source luminosity between 0.3--3.5 keV in units of
$10^{44}$ erg s$^{-1}$, $\gamma_{1}=1.7$,
$\gamma_{2}=3.4$, while 
$K_1 (=K_2L_B^{\gamma1-\gamma2}$) and $L_B$ now take the values 
$6.4\times10^{-7}$ Mpc$^{-3}\,(10^{44}$ erg s$^{-1})^{\gamma_{1}-1}$ and
$8.1\times10^{43}$  erg s$^{-1}$, respectively.

 The XLF, which spans the luminosity range $10^{42}-10^{47}$ erg
s$^{-1}$,
has been assumed to evolve up to $z_{cut}=1.8$ and to be constant
between $z_{cut}$ and $z_{max}=3$. The input X--ray spectra are those
described by Comastri et al. (1995).
The absorbed AGN are divided into four bins
centered at log$N_{\rm H}=21.5, 22.5, 23.5, 24.5$, to cover the $N_{\rm
H}$ range $10^{21}-10^{25}$ cm$^{-2}$. Their number densities, which are
free parameters of the model, normalized 
to the density of unabsorbed AGN are
respectively: 0.35,1.30, 2.0, 1.4.
The iron lines have been subsequently added in the input spectra.  

Extremely absorbed AGN ($N_{\rm
H}>10^{25}$ cm$^{-2}$) are not expected to contribute significantly
to the XRB ( Comastri et al. 1995). Indeed, their 2--10 keV
luminosity, entirely due to the reflection
continuum, is found to be two order of magnitude lower than 
that of Sey 1s in the same band (Maiolino et al. 1998).
However, given their very strong lines, we have included them
in our model calculation. 
We have then assumed as the shape of the XLF of these objects that of
type 1 AGN, downshifted in flux by a factor 100, and
as their typical spectrum a pure reflection continuum (e.g.
Lightman \& White 1988) normalized to reproduce the 
observed luminosities.
The number density of these sources is very uncertain. Preliminary
estimates (Maiolino et al. 1998; Risaliti et al. 1998) 
indicate that the density of Compton thick AGN with 
$N_{\rm H}>10^{25}$ cm$^{-2}$ is 15--20\% of the density of all type 2s.
If we consider as type 2s the objects with $N_{\rm H}>10^{22}$
cm$^{-2}$ (see Fig.~1 and 2), the normalized density of AGN with $N_{\rm 
H}>10^{25}$ cm$^{-2}$ in our model results 0.8.
As it will be shown in the next section, the large uncertainty on
this value does not have relevant effects in our model.

In agreement with the observed X--ray spectra of quasars, we have not
added the iron line in type 1 objects with luminosity above $L_B$.

We have chosen a simple gaussian shape for the
line profile, which is sufficiently
accurate to estimate the line contribution to the XRB, the most
important parameter being the line equivalent width. 
As a first order approximation, we have added to the objects 
with $N_{\rm H}<10^{24}$ cm$^{-2}$ a line with constant $W_{\rm
K\alpha}$,
regardless of the amount of intrinsic absorption. 
In order to maximize the effect produced by the iron line we 
have performed calculations with $W_{\rm K\alpha}=390$ eV, 
which corresponds to the maximum value quoted in Table~2. 
For the objects with larger $N_{\rm H}$ we have consistently added a
line with $W_{\rm K\alpha}$=2 keV.
Calculations have been carried out for both narrow 
($\sigma_{\rm K\alpha}=0.1$ keV) and broad ($\sigma_{\rm K\alpha}=0.4$
keV) lines. 
In addition we have run a model by adopting the average
$W_{\rm K\alpha}$ values pertaining to the different bins of
log$N_{\rm H}$ as shown in Table~4.

\begin{table}[h]
\caption{Mean equivalent widths per log$N_{\rm H}$ interval from the ASCA
sample.}
\begin{tabular}{cccccc}
\hline \hline
log[$N_{\rm H}$(cm$^{-2}$)]& $<21$& 21.5& 22.5& 23.5& $>24$\\
\hline
$W_{\rm K\alpha}$(eV)& 280& 370& 300& 470& 2100\cr
\hline
\end{tabular}
\end{table}

\section{Results}

\begin{figure}
\epsfig{file=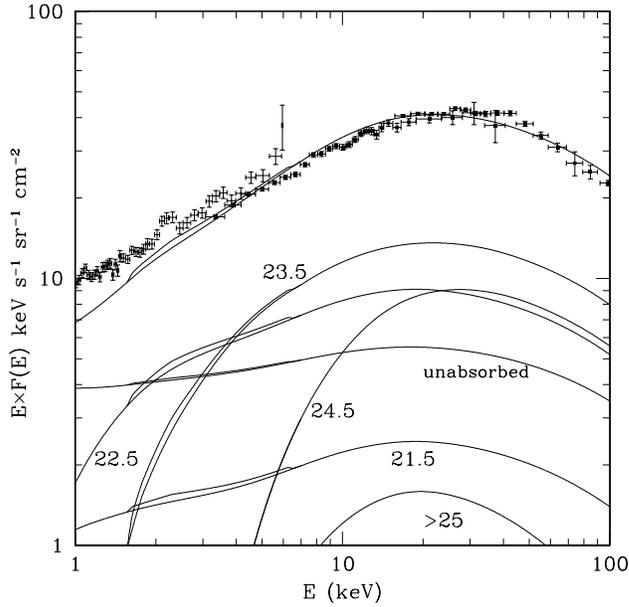, height=9.cm, angle=0, silent=}
\caption{The fit to the XRB. The data of larger intensity below 6 keV are
from ASCA (Gendreau
et al. 1995), while those above 3 keV are a compendium of
various
experiments including HEAO--1 A2 (Gruber 1992). The various curves
represent
the contributions of the different classes of sources, the labels
being
the logarithms of the hydrogen column densities.
The curve of unabsorbed
sources includes both Sey 1s and quasars. The thick lines between
$1.5-7$ keV represent the additional fluxes due to the iron emission
lines
for $W_{\rm K\alpha}=390$ eV and $\sigma_{\rm K\alpha}=0.1$ keV.}
\end{figure}

The model fit to the XRB spectrum is shown in Fig.~3. This is based
on the
data above 3 keV derived from the HEAO--1 A2 experiment, thus it does
not match with the ASCA measurements which have a higher normalization
($10-15\%$ at 3 keV). 
However, since the iron signatures are expected in
the ASCA band,  we will also compare our results with the
the Gendreau et al. (1995) data (see below).

The labeled curves represent the contribution of the different absorption
classes to the overall XRB, the labels being the logarithm of the hydrogen
column density. The curve of unabsorbed objects includes both Sey 1s and 
quasars.
For each curve the thick line between $1.5-7$ keV represents
the additional flux due to the iron line. Note that for
objects with log$N_{\rm H}=24.5$ the line excess is
undistinguishable from the underlying continuum. This is due to the 
extremely hard shape of the highly absorbed sources.
As expected, the
contribution of AGN with log$N_{\rm H}>25$ to the XRB is very small:
below 4\% at about 20 keV and negligible in the 1--7 keV band.

\begin{figure}
\epsfig{file=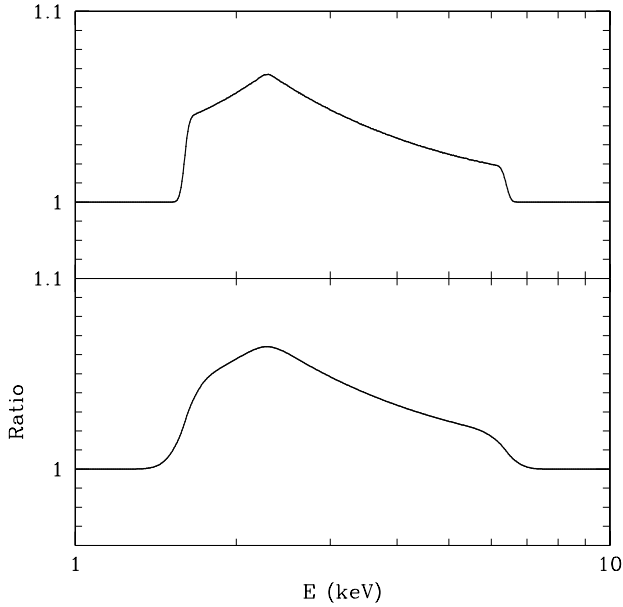, height=9.cm, angle=0, silent=}
\caption{The iron line contribution normalized to the XRB model continuum.
The curves in Panels a) and b) correspond
respectively to $\sigma_{\rm K\alpha}=0.1$ keV and
$\sigma_{\rm K\alpha}=0.4$ keV for $W_{\rm K\alpha}=390$ eV.}
\end{figure}

An enlarged view of the relative
contribution of the iron line emission with respect to the model without the
line is given in Fig.~4.
The iron line profile, smeared out by the redshift effects, is
relatively smooth: the maximum attained level with
respect to the model without the
line is less than 7\% at $6.4/(1+z_{cut})\simeq2.3$ keV 
for $W_{\rm K\alpha}=390$ eV. 
Once the $W_{\rm K\alpha}$ value has been fixed, it is found, as
expected, that
the only effect produced by a broader line is a smoother profile.
When using the $W_{\rm K\alpha}$ values in Table ~4, and adopting
$\sigma_{\rm K\alpha}=0.4$ keV when log$N_{\rm H}< 22$ and
$\sigma_{\rm K\alpha}=0.1$ keV when log$N_{\rm H}\geq 22$, 
the maximum line contribution is still below 7\%.

Matt et al. (1996) have suggested that in Compton thick objects, together
with a strong fluorescent iron line at 6.4 keV, another strong iron line 
at 6.7
keV, produced by resonant scattering in the warm mirror, should be present.
Both these lines are predicted to have equivalent widths of a few keV. 
Indeed, strong emission lines from highly ionized iron 
with $W_{\rm K\alpha}$ values above 1 keV have been detected in 
some of the Compton thick objects observed so far, such as NGC 1068 
(Ueno et al. 1994) and NGC 2273 (Maiolino et al. 1998). 
We estimate that a
line with $W_{\rm K\alpha}=5$ keV is sufficiently large to
approximate the
effects of the iron lines complex of highly absorbed AGN. By performing
this test we find that the additional iron flux 
is less than 1\%.

\begin{figure}
\epsfig{file=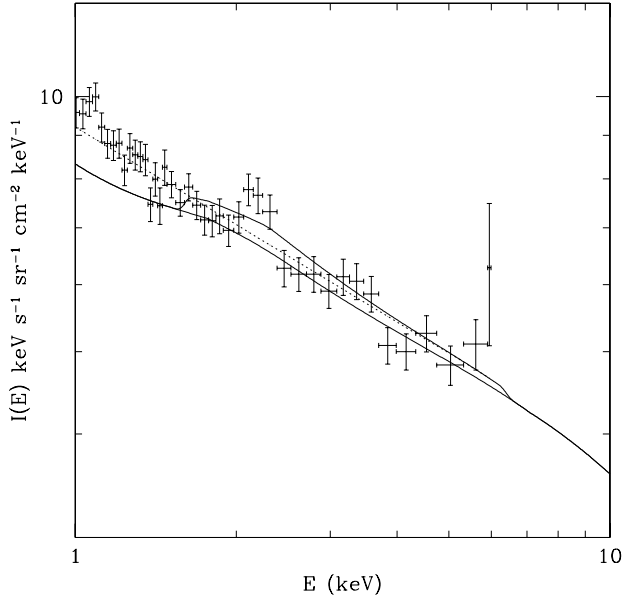, height=9.cm, angle=0, silent=}
\caption{The spectrum of the 1--6 keV ASCA XRB (Gendreau et al. 1995)
compared with our model with/without the line contribution (upper/lower
solid
line) for $W_{\rm K\alpha}$=390 eV and $\sigma_{\rm K\alpha}$=0.1
keV.
The dotted line is the best power law fit to the XRB.}
\end{figure}

Since we are mainly interested on the effects of the iron feature 
on the shape of the XRB spectrum we have rescaled 
the model in order to match the XRB intensity observed by ASCA.
The comparison between the rescaled model and the ASCA data is shown 
in Fig.~5. The data are the same as given by Gendreau et al. (1995) but
with a different binning.
It should be noted that 
the model does not reproduce the  
data below $\sim 2$ keV, because in this spectral region the contribution
of sources other than AGN is expected (Comastri el al. 1995).
Clearly, the additional flux provided
by the iron line is below present
uncertainties in the XRB measurements and the dispersion of the data does
not allow yet to detect the predicted iron feature.
It should also be noted that the relatively deep
gold edge in the instrumental response at $\sim2.2$ keV, which introduces
calibration uncertainties around this energy, is very close to the energy
where the maximum deviations due to the iron line are expected.
Therefore the narrow bump in the observed                 
XRB spectrum around 2 keV 
has an instrumental origin and is not produced by the iron line. 

\section{Conclusions}

The main conclusion of this work is
that the predicted contribution of the fluorescence iron lines to the XRB
in the 1.5 -- 7 keV energy interval is $\stackrel{<}{_{\sim}} 5\%$ 
with a
peak intensity $\stackrel{<}{_{\sim}} 7\%$
at $\sim 2$ keV. Although in our computations we have adopted 
the specific model of Comastri et al. (1995), this result should essentially
hold for those synthesis models of the XRB that are based on the AGN
unifications schemes.
This estimated contribution of the iron lines is still undetectable within 
the dispersion of the best available data. 
Indeed, the uncertainties in the power law fit
of the XRB spectrum
observed by ASCA  are of the order of $\sim10\%$ (Gendreau et al. 1995).
In order to reproduce an iron excess of more than $\sim10\%$ 
a mean equivalent width of $W_{\rm K\alpha}\stackrel{>}{_{\sim}}600$ eV
for the
sources with $N_{\rm H}<10^{24}$ cm$^{-2}$ should have been adopted, which
is
inconsistent with
the observed average values.

Since in our calculations we have assigned to all Sey nuclei an
iron line equivalent width of 390 eV, corresponding to the maximum average
value quoted in Table~2, and since not all observed nuclei show
detectable iron
lines, we are confident that our result may represent a rather 
stringent upper limit to any iron line contribution to the XRB.
Therefore, the present undetection of iron features in the XRB
spectrum does not constitute by itself
an argument against the synthesis of the XRB by known classes of AGN.
On the contrary, one may argue that the 
detection of the iron signature would constitute strong evidence in
favour of these models, while putting at the same time severe constraints
on other classes of as yet undetected AGN which have been
recently hypothesized in order to explain the XRB.

In order to observe the predicted iron feature,
the dispersion in the XRB data should be reduced by at least 
50\%. Perhaps, this may be achieved in the near future with the forthcoming
X--ray missions. It may be noted that,
in principle, from the energy range and profile of the iron feature
one could also get relevant information about the  
values of $z_{max}$ and $z_{cut}$ characterizing the evolution of AGN.

Finally, we note that the estimated intensity of the iron feature
may partially explain the
discrepancy in the normalizations of ASCA and HEAO--1 A2 data. Since the
normalization of the XRB from HEAO--1 A2 has been derived from a broad
band analysis of the data, one may wonder whether the intensity at
the low energy end of
the spectrum, where the iron feature is expected, has been somewhat
underestimated.

\begin{ack}

We thank K. Gendreau for providing the data on the ASCA XRB.
AC acknowledges financial support from the Italian Space Agency 
under contract ARS--96--70. RG acknowledges financial support from the
Italian Space Agency under contract ARS--98--116/22.

\end{ack}

\end{document}